\documentclass[aps, superscriptaddress, twocolumn, nopacs, nobibnotes, floatfix, amsmath, amssymb, letterpaper]{revtex4-1}
\setcitestyle{super}
\usepackage[utf8]{inputenc}
\usepackage[T1]{fontenc}
\usepackage{graphicx}
\usepackage{amsmath, amssymb}
\usepackage{hyperref}
\usepackage{chngcntr}
\usepackage{xcolor}

\begin{document}

\title{Nature of the current-induced insulator-to-metal transition\\ in Ca$_2$RuO$_4$ as revealed by transport-ARPES}

\author{C. T. \surname{Suen}}
\email{c.suen@fkf.mpg.de}
\affiliation{Max Planck Institute for Solid State Research, Stuttgart, 70569, Germany}
\affiliation{Quantum Matter Institute, University of British Columbia, Vancouver, V6T 1Z4, Canada}
\affiliation{Department of Physics and Astronomy, University of British Columbia, Vancouver, V6T 1Z1, Canada}
\affiliation{Advanced Light Source, Lawrence Berkeley National Lab, Berkeley, 94720, USA}

\author{I. \surname{Markovi\'{c}}}
\affiliation{Quantum Matter Institute, University of British Columbia, Vancouver, V6T 1Z4, Canada}
\affiliation{Department of Physics and Astronomy, University of British Columbia, Vancouver, V6T 1Z1, Canada}
\affiliation{Canadian Light Source, 44 Innovation Boulevard, Saskatoon, S7N 2V3, Canada}

\author{M. \surname{Zonno}}
\affiliation{Canadian Light Source, 44 Innovation Boulevard, Saskatoon, S7N 2V3, Canada}

\author{N. \surname{Heinsdorf}}
\affiliation{Max Planck Institute for Solid State Research, Stuttgart, 70569, Germany}
\affiliation{Quantum Matter Institute, University of British Columbia, Vancouver, V6T 1Z4, Canada}
\affiliation{Department of Physics and Astronomy, University of British Columbia, Vancouver, V6T 1Z1, Canada}

\author{S. \surname{Zhdanovich}}
\affiliation{Quantum Matter Institute, University of British Columbia, Vancouver, V6T 1Z4, Canada}
\affiliation{Department of Physics and Astronomy, University of British Columbia, Vancouver, V6T 1Z1, Canada}

\author{N.-H. \surname{Jo}}
\affiliation{Advanced Light Source, Lawrence Berkeley National Lab, Berkeley, 94720, USA}

\author{M. \surname{Schmid}}
\affiliation{Max Planck Institute for Solid State Research, Stuttgart, 70569, Germany}

\author{P. \surname{Hansmann}}
\affiliation{Friedrich-Alexander-Universit\"{a}t Erlangen-N\"{u}rnberg, Erlangen, Germany}

\author{P. \surname{Puphal}}
\affiliation{Max Planck Institute for Solid State Research, Stuttgart, 70569, Germany}

\author{K. \surname{F\"{u}rsich}}
\affiliation{Max Planck Institute for Solid State Research, Stuttgart, 70569, Germany}

\author{V. \surname{Zimmermann}}
\affiliation{Max Planck Institute for Solid State Research, Stuttgart, 70569, Germany}
\affiliation{Quantum Matter Institute, University of British Columbia, Vancouver, V6T 1Z4, Canada}
\affiliation{Department of Physics and Astronomy, University of British Columbia, Vancouver, V6T 1Z1, Canada}

\author{S. \surname{Smit}}
\affiliation{Quantum Matter Institute, University of British Columbia, Vancouver, V6T 1Z4, Canada}
\affiliation{Department of Physics and Astronomy, University of British Columbia, Vancouver, V6T 1Z1, Canada}

\author{C. \surname{Au-Yeung}}
\affiliation{Quantum Matter Institute, University of British Columbia, Vancouver, V6T 1Z4, Canada}
\affiliation{Department of Physics and Astronomy, University of British Columbia, Vancouver, V6T 1Z1, Canada}

\author{B. \surname{Zwartsenberg}}
\affiliation{Quantum Matter Institute, University of British Columbia, Vancouver, V6T 1Z4, Canada}
\affiliation{Department of Physics and Astronomy, University of British Columbia, Vancouver, V6T 1Z1, Canada}

\author{M. \surname{Krautloher}}
\affiliation{Max Planck Institute for Solid State Research, Stuttgart, 70569, Germany}

\author{I. S. \surname{Elfimov}}
\affiliation{Quantum Matter Institute, University of British Columbia, Vancouver, V6T 1Z4, Canada}
\affiliation{Department of Physics and Astronomy, University of British Columbia, Vancouver, V6T 1Z1, Canada}

\author{R. \surname{Koch}}
\affiliation{Advanced Light Source, Lawrence Berkeley National Lab, Berkeley, 94720, USA}

\author{S. \surname{Gorovikov}}
\affiliation{Canadian Light Source, 44 Innovation Boulevard, Saskatoon, S7N 2V3, Canada}

\author{C. \surname{Jozwiak}}
\affiliation{Advanced Light Source, Lawrence Berkeley National Lab, Berkeley, 94720, USA}

\author{A. \surname{Bostwick}}
\affiliation{Advanced Light Source, Lawrence Berkeley National Lab, Berkeley, 94720, USA}

\author{M. \surname{Franz}}
\affiliation{Quantum Matter Institute, University of British Columbia, Vancouver, V6T 1Z4, Canada}
\affiliation{Department of Physics and Astronomy, University of British Columbia, Vancouver, V6T 1Z1, Canada}

\author{E. \surname{Rotenberg}}
\affiliation{Advanced Light Source, Lawrence Berkeley National Lab, Berkeley, 94720, USA}

\author{B. \surname{Keimer}}
\email{b.keimer@fkf.mpg.de}
\affiliation{Max Planck Institute for Solid State Research, Stuttgart, 70569, Germany}

\author{A. \surname{Damascelli}}
\email{damascelli@physics.ubc.ca}
\affiliation{Quantum Matter Institute, University of British Columbia, Vancouver, V6T 1Z4, Canada}
\affiliation{Department of Physics and Astronomy, University of British Columbia, Vancouver, V6T 1Z1, Canada}

\date{\today}


\maketitle

\textbf{The Mott insulator Ca$_2$RuO$_4$ exhibits a rare insulator-to-metal transition (IMT) induced by DC current \cite{okazaki2013}. While structural changes associated with this transition have been tracked by neutron diffraction, Raman scattering, and x-ray spectroscopy, work on elucidating the response of the electronic degrees of freedom is still in progress \cite{bertinshaw2019, fuersich2019, jenni2020, curcio2023, zhao2019, sakaki2013}. Here we unveil the current-induced modifications of the electronic states of Ca$_2$RuO$_4$ by employing angle-resolved photoemission spectroscopy (ARPES) in conjunction with four-probe transport. Two main effects emerge: a clear reduction of the Mott gap and a modification in the dispersion of the Ru-bands. The changes in dispersion occur exclusively along the $XM$ high-symmetry direction, parallel to the $b$-axis where the greatest in-plane lattice change occurs. These experimental observations, together with dynamical mean-field theory (DMFT) calculations simulated from the current-induced structural distortions, indicate the intimate interplay of lattice and orbital-dependent electronic response in the current-driven IMT. Furthermore, based on a free energy analysis, we demonstrate that the current-induced phase, albeit thermodynamically equivalent, is electronically distinct from the high-temperature zero-current metallic phase. Our results provide insight into the elusive nature of the current-induced IMT of Ca$_2$RuO$_4$ and advance the challenging, yet powerful, technique of transport-ARPES.}

Ca$_2$RuO$_4$ (CRO) is a quasi-two-dimensional system comprising of layers of RuO$_6$ octahedra that undergo an insulator-to-metal transition (IMT) when heated above $T_{\mathrm{IMT}} = 357$~K \cite{braden1998, alexander1999}. The low temperature insulating (high temperature metallic) phase is commonly referred to as S phase (L phase), due to the shorter (longer) $c$-axis. At low temperatures, the $c$-axis compression tilts and distorts the RuO$_6$ octahedra, lifting the electron degeneracy of the Ru $t_{2g}$ states, and inducing a Mott-insulating gap~\cite{gorelov2010, sutter2017}. This sensitivity of the ground state to the octahedral compression renders the system extremely prone to external perturbation. Aside from temperature, the IMT can also be tuned through chemical substitution \cite{nakatsuji2000, friedt2001}, pressure \cite{nakamura2002, steffens2005}, strain \cite{dietl2018, ricco2018}, electric fields \cite{nakamura2013}, and interestingly, the aforementioned DC current \cite{okazaki2013}. In the latter case, the metallic phase persists so long as current is continuously passed through the sample, making it a rare example of a long-lasting steady state in a field of non-equilibrium physics dominated by pump-probe techniques \cite{boschini2023, claudioRMP}.

While the zero-current IMT is understood as a consequence of the shortened $c$-axis, the mechanism by which the current induces the IMT is not established. The current-induced structural changes are distinct from those driven by temperature, with the insulating (metallic) state exhibiting a slightly larger (smaller) $c$-axis compared to the zero-current counterparts, earning the S* and L* denominations, respectively. The current also suppresses the antiferromagnetic order that typically occurs below $T_{\mathrm{N}}$ = 110~K. While this IMT is observed regardless of whether the current is applied within the $ab$-plane \cite{fuersich2019, mattoni2020, zhang2019} or along the $c$-axis \cite{bertinshaw2019}, an unusual metal-insulator nanostripe pattern arises when intermediate current densities are applied exclusively parallel to the $b$-axis of the crystal \cite{zhang2019}.

\begin{figure*}[t]%
\centering
\includegraphics[width=0.999\textwidth]{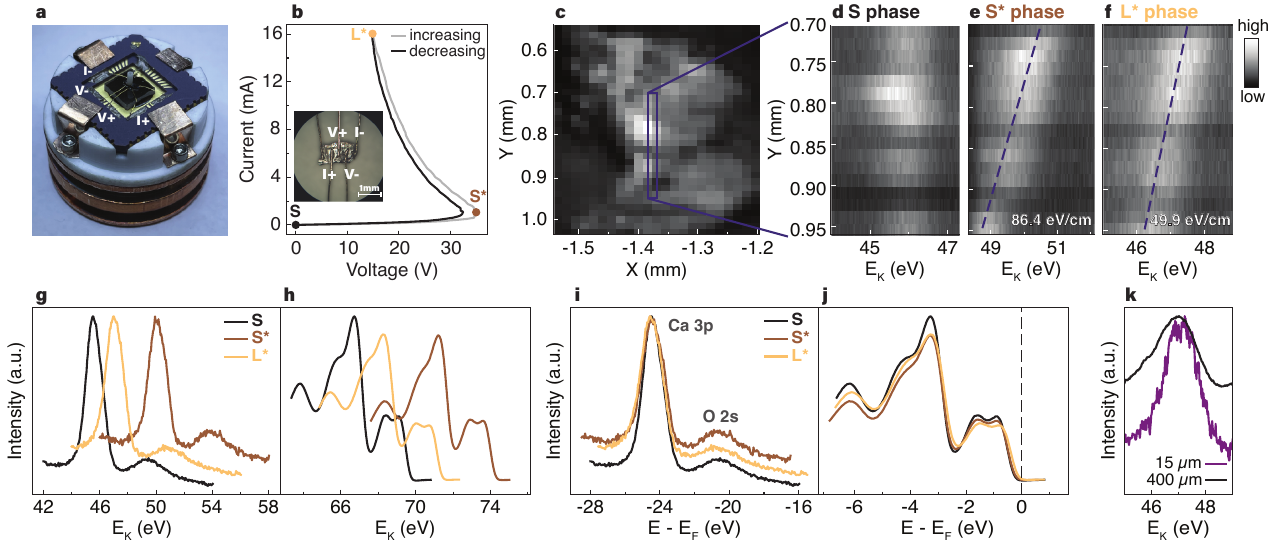}
\caption{\textbf{ARPES under current.} (a) Experimental configuration for transport-ARPES. (b) $I$-$V$ curve taken during the transport-ARPES experiment showing the negative differential resistance characteristic of the current-induced IMT. Inset shows contacts on the back side of the sample. (c) Spatial map of the cleaved section of the sample, measured by moving a 15~\textmu m spot size across the surface in zero-current conditions. False colour represents the peak intensity of the Ca $3p$ core level. (d - f) Variation of the Ca $3p$ core level energy with sample position, along the direction of current flow ($y$-axis); position range is indicated by the purple box in (c). (g - j) Correction of the energy shifts in the spectra: (g) Ca $3p$ core levels and (h) valence bands as measured for the three structurally distinct phases at ($x, y$) = (-1.37, 0.75). (i, j) The same data as (g, h) after correction for the energy shift by alignment of the Ca $3p$ core levels to that of the S phase. All spectra are taken at 74~eV with $\sigma$-polarized light. (k) Effect of broadening on the Ca $3p$ core level due to the electric potential present across a 15~\textmu m (purple) and 400~$\mu$m (black) diameter beamspot: core levels displayed are integrated over the specified diameters from data shown in (f); in addition to the broadening of the 400~$\mu$m curve, the shift in energy induced by the electric field can also be observed.}
\label{fig1}
\end{figure*}

In order to pinpoint the mechanism behind the current-driven IMT in CRO, an understanding of the electronic structure evolution between the insulating and current-driven steady states are essential. Here, we combine transport with angle-resolved photoemission spectroscopy (transport-ARPES) to directly probe the electronic structure under current. Both transport and ARPES probe scattering processes and density of states; however, transport is an integrative probe over the whole sample, whereas ARPES can probe the state of the system point-by-point within the spatial resolution of the beamspot. The merging of an ‘integrative’ with a ‘scanning’ probe allows us to better differentiate between crystal defects and intrinsic microscale phenomena. Although transport-ARPES has the potential to study the effects of current on phase transitions~\cite{tsen2015}, superconductivity~\cite{goren2010, kaminski2016}, charge density waves~\cite{lee1979}, and competing orders in strongly correlated systems~\cite{cao2018}, its wide application has, to date, been precluded by the complexity of dissociating real current-driven changes from the effects of the stray electric and magnetic fields generated by the current \cite{kaminski2016, curcio2020, majchrzak2021, hofmann2021, curcio2023}.

\begin{figure*}[t]%
\centering
\includegraphics[width=0.999\textwidth]{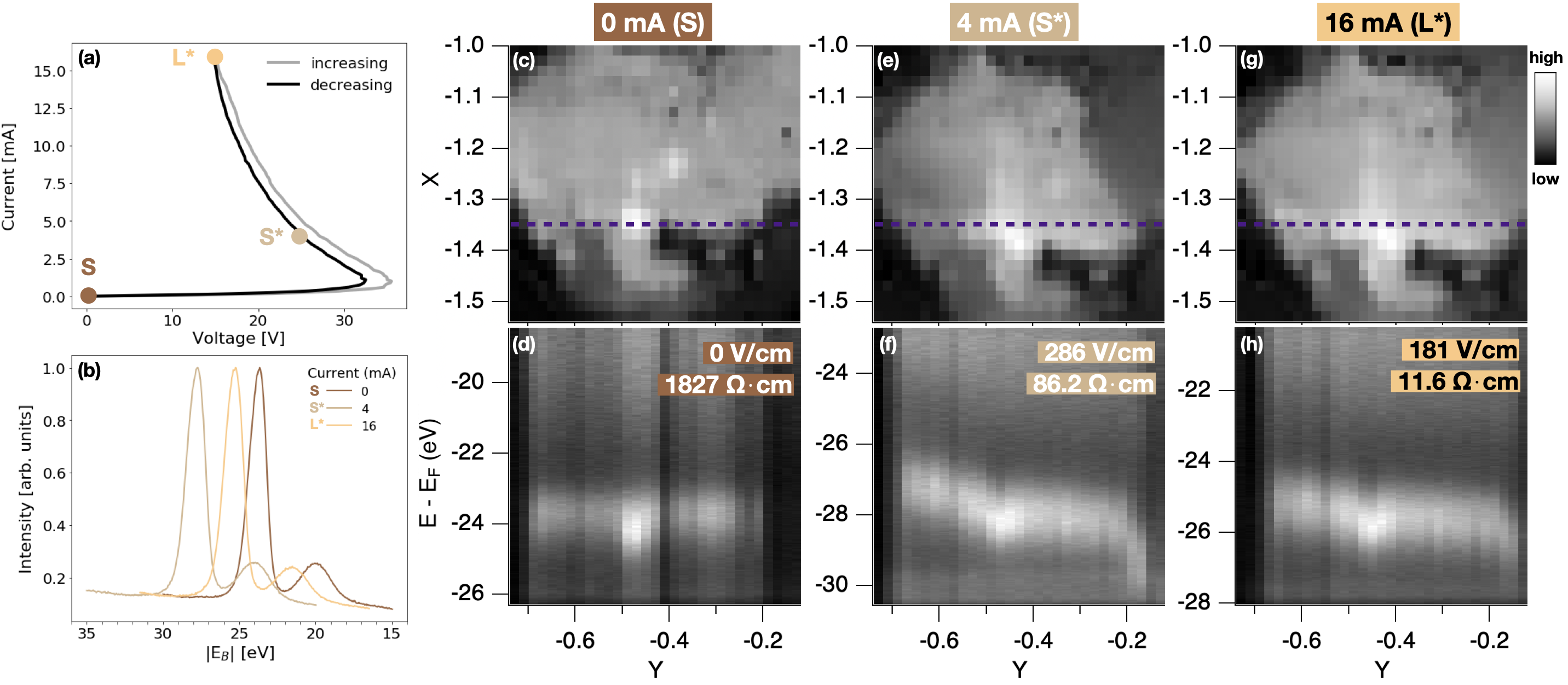}
\caption{\textbf{Charge-gap reduction through the IMT.} (a) Experimental Fermi surface in the L* phase, symmetrized and integrated over $E_{\mathrm{F}}=\pm 50$~meV; the orange square represents the boundary of the first Brillouin zone, with the letters denoting points of high symmetry with respect to the orthorhombic Brillouin zone. (b) Fermi surface of CRO in the L* phase from DMFT calculations. (c) Total measured ARPES spectral weight, integrated over the entire Brillouin zone in the S (black) and L* phases (yellow), showing an overall charge-gap reduction. (d) Spectral function of CRO in the S (black) and L* phases (yellow) integrated over the Brillouin zone from DMFT calculations. (e) Measured ARPES dispersions along the $XM$ high-symmetry direction without (left) and with (right) applied current. Ru bands are found above $-2.5$~eV and O bands below $-2.5$~eV. (f) ARPES spectral weight integrated along the $XM$ direction for the S, S* and L* phases (see Fig.~\ref{fig1}b for definitions). The S phase dashed line represents the spectral weight after the current has been ramped down to 0~mA.}
\label{fig2}
\end{figure*}

\begin{figure*}[t]%
\centering
\includegraphics[width=0.999\textwidth]{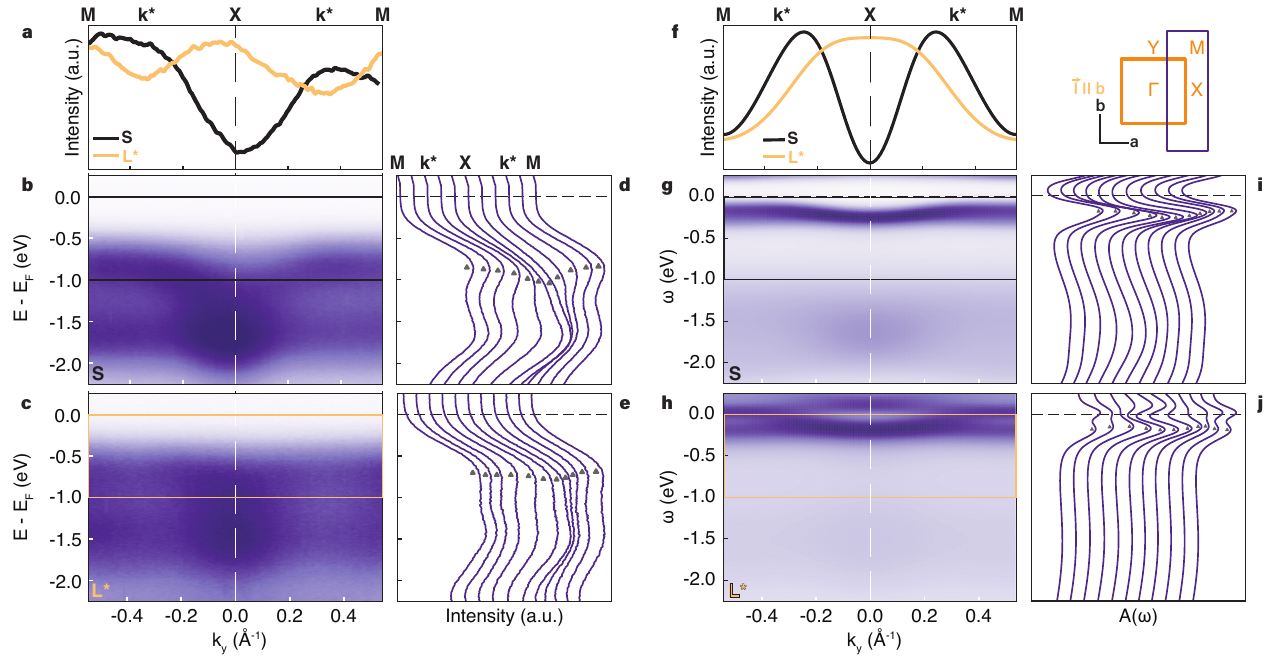}
\caption{\textbf{Change in the Ru-band dispersion along $MXM$.} (a) MDCs integrated between -1~eV to $E_{\mathrm{F}}$ as highlighted by the black (S) and yellow (L*) boxes in the energy vs. momentum plots for the (b) S and (c) L* phases. $k^*$ marks the momentum where spectral weight is observed to decrease with current. (d, e) EDCs integrated over 0.1~\AA$^{-1}$ across the valence bands along the $MXM$ direction, in the (d) S and (e) L* phases [see also schematic above (i) which displays the experimental configuration]. Gray triangles guide the eye along the progression of the lower energy Ru peak across momentum. (f) MDCs integrated between -1~eV to $E_{\mathrm{F}}$ as highlighted by the black (S) and yellow (L*) boxes in the DMFT spectral function calculations for the (g) S and (h) L* phases. Note, the spectrum is significantly sharper in the vicinity of $E_{\mathrm{F}}$; however, there is also sizeable spectral weight around -1.5 to -2.0~eV, in the region of the Hubbard band (see Fig.~\ref{fig2}d). A Gaussian convolution has been applied to the data for better comparison to experiment. (i, j) EDCs integrated over 0.1~\AA$^{-1}$ across the valence bands from along the $MXM$ direction, in the (i) S and (j) L* phases.}
\label{fig3}
\end{figure*}

\begin{figure*}[t]%
\centering
\includegraphics[width=0.999\textwidth]{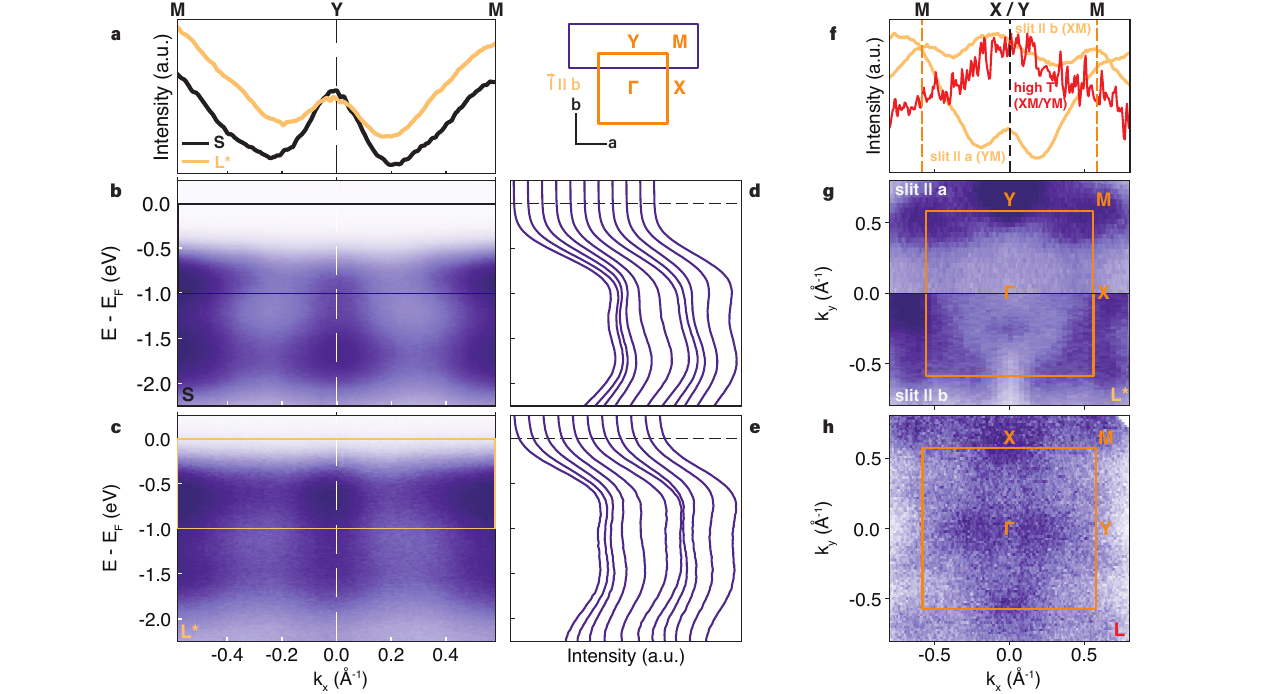}
\caption{\textbf{Ru bands along $MYM$ and high temperature data.} (a) MDCs integrated between -1~eV to $E_{\mathrm{F}}$ as highlighted by the black (S) and yellow (L*) boxes in the energy vs. momentum plots for the (b) S and (c) L* phases [see also schematic above (d) highlighting the experimental configuration]. (d, e) EDCs integrated over 0.1~\AA$^{-1}$ across the valence bands from along the $MYM$ direction, in the (d) S and (e) L* phases. (f) MDCs integrated over 0.5~eV for the L* phase along $XM$ and $YM$ (yellow) cuts, compared to the high temperature L phase (red) cut. Note, the L phase MDC is labelled as both $XM$ and $YM$ because the same Fermi surface is measured when either axes is aligned parallel to the slit (see Supplementary section F. Additionally, the yellow L* curves correspond to the curves found in Fig.~\ref{fig3}a and Fig.~\ref{fig4}a. (g) Fermi surface in the L* phase measured with the analyser slit oriented along the $a$-axis of the sample (top) and the $b$-axis of the sample (bottom), the latter of which is shown in full in Fig.~\ref{fig2}a. (h) Zero-current L phase Fermi surface measured at $T = 384$~K, i.e. above $T_{\mathrm{IMT}} = 357$~K.}
\label{fig4}
\end{figure*}

We mitigate these challenges by using a micron-size beamspot in conjunction with core-level spectroscopy to establish a common binding energy reference at different magnitudes of applied current (Fig.~\ref{fig1}). This approach uncovers clear signatures of the current-induced IMT in the electronic structure of CRO: a global reduction of the charge gap (Fig.~\ref{fig2}) and a change in the low-energy Ru-bands at the edge of the Brillouin zone (Fig.~\ref{fig3}). These changes in dispersion occur exclusively parallel to the axis that undergoes a larger structural modification under current, suggesting that the current prompts intimately intertwined electronic and structural changes, which are the main driver of the IMT. Additionally, we find the resulting Fermi surface to be distinct from that of the high temperature phase, indicating that the current-induced phase is not simply a consequence of Joule heating (Fig.~\ref{fig4}h).



Our transport-ARPES  experimental configuration is presented in Fig.~\ref{fig1}a, along with the characteristic $I$-$V$ curve in Fig.~\ref{fig1}b. The distinctive negative differential resistance of the current-induced transition~\cite{zhang2019, jenni2020} indicates a high quality sample devoid of large cracks or deformations that would globally impede current flow~\cite{mattoni2020}. To avoid possible domain formation, such as the metal-insulator nanostripe pattern found at intermediate current densities \cite{zhang2019}, we refer to the current-voltage characteristic to ensure that the magnitude of the applied current is large enough such that the whole sample has completely transitioned into the L* phase. This critical current will depend on the dimensions of the particular sample and the temperature at which the experiment is carried out \cite{mattoni2020}.

Before presenting the transport-ARPES results on CRO, we briefly discuss our experimental strategy to address the effects of the stray electric and magnetic fields generated by an electric current passed through the sample (Supplementary section C). The magnetic field results in a rigid momentum shift and deformation of the measured band structure; the latter has minimal impact in our experiment, whereas the former is accounted for by setting the high symmetry points of the measured spectra to their established positions in \textit{\textbf{k}}-space (Fig.~C.1). On the other hand, the potential gradient that develops across the sample as a consequence of current flow causes a deviation in energy and broadens spectral features; this gradient shifts the kinetic energy of emitted photoelectrons by the magnitude of the local potential at the emission position \cite{bostwick2009}. We illustrate this in Figs.~\ref{fig1}(d - f) by tracking the Ca $3p$ core level binding energy: the peak shifts as the beam is physically moved along the direction of current flow. Note that the S* phase exhibits a larger shift, demonstrating that it experiences a larger local electric field, than the L* phase at higher current [Figs.~\ref{fig1}(e, f)], consistent with the negative differential resistance of the current-induced IMT. The slope of the spatial variation of the core level illustrates the average field present across that particular sample region. The presence of smooth slopes indicates which regions have homogeneous current flow suitable for ARPES measurement, in contrast to regions compromised by impurities or cracks within the crystal.

While changes in the low-energy electronic states are expected to occur due to the IMT, core levels will be significantly less affected. As neither the core level lineshapes, nor the Ca $3p$-O $2s$ energy difference, change evidently across the IMT (Fig.~\ref{fig1}i), we conclude that the dominant cause of the energy shift is the electrostatic potential, rather than chemical shifts, within a small uncertainty of order $\sim 100$~meV [Supplementary section C]. Such small chemical shifts would not affect the main conclusions discussed below. This allows us to apply the extracted core level shifts of 4.5~eV between the S and S* phases, and 1.6~eV between the S and L* phases, to the valence band spectra obtained at the same spot on the sample to expose the intrinsic current-induced modifications close to the chemical potential (Fig.~\ref{fig1}h, j).

In addition, since photoemission data are simultaneously acquired over the area illuminated by a finite sized beam spot, photoelectrons experience a range of energy shifts given by the potential gradient across the spot, as seen in (Fig.~\ref{fig1}k). To minimize this range and enable the observation of band shifts on the order of the insulating gap in CRO ($\sim 0.4$~eV \cite{nakatsuji2000}), we use a 15~$\mu$m diameter spot. This results in a current-induced broadening of $\Delta E_{\mathrm{k}} \sim 75$~meV (Supplementary section C), which allows us to probe the intrinsic L* phase electronic signatures discussed in the following sections. Once the effects of the current-induced stray fields are accounted for, we uncover the IMT-driven modifications of the valence band as the system transitions from the S to L* phase. Note that, despite the crystal structure decreasing in orthorhombicity as it goes through the IMT, high symmetry points are denoted with respect to the orthorhombic Brillouin zone.

Fermi surfaces in the current-induced metallic L* phase obtained from ARPES and DMFT are presented in Figs.~\ref{fig2}a and b. While the charge gap removes spectral weight at the Fermi level in the insulating S phase, with applied current we clearly detect momentum dependent spectral weight at $E_{\mathrm{F}}$ (Fig.~\ref{fig2}a). Indeed, a reduction of the total gap in CRO emerges when the spectral weight integrated over the whole Brillouin zone between the S and L* phases is compared (Fig.~\ref{fig2}c), as also recently observed by Curcio et. al~\cite{curcio2023}. This is further illustrated by the ARPES maps acquired along $XM$ in the two phases (Fig.~\ref{fig2}e), and by the evolution of the energy distribution curves (EDCs) integrated over the orthorhombic Brillouin zone edge for zero, intermediate, and high current (Fig.~\ref{fig2}f). We emphasize that the gap reduction occurs only when the IMT is fully crossed (brown to yellow curves) and is accompanied by a redistribution of the spectral weight towards lower binding energies for the whole Ru and O valence band manifold. These observations are in good agreement with the metallic Fermi surface and momentum integrated spectral function obtained by DMFT calculations (Figs.~\ref{fig2}b, d), in which the current is accounted for through the use of the current-induced lattice constants obtained by neutron diffraction~\cite{bertinshaw2019}. 



While no noticeable differences are observed in the dispersion of the O-bands (Fig.~\ref{fig2}e), the $XM$ dispersion of the Ru-bands undergoes a distinct change through the IMT (Fig.~\ref{fig3}). 
Without current, flat bands around $k^* = \pm 0.38$~\AA$^{-1}$ are observed to disperse downwards into the $X$ point, and all bands are found well below $E_{\mathrm{F}}$, consistent with the insulating phase (Fig.~\ref{fig3}b). Upon applying current, two main changes occur (Fig.~\ref{fig3}c): the spectral weight at the $X$ point close to the Fermi level increases, whereas the spectral weight at $k^*$ decreases. These effects are further illustrated by the momentum distribution curves (MDCs) shown in Fig.~\ref{fig3}a. While a trough at $X$ is enveloped by two peaks at $k^*$ in the zero-current MDC, a peak is enveloped by two troughs in the current-induced MDC. Similarly, while the cascade of EDCs in Fig.~\ref{fig3}d exhibits a clear modulation of spectral weight across the $MXM$ cut, the EDCs of the L* phase in Fig.~\ref{fig3}e are fairly momentum independent. These changes were not observed by Curcio et. al~\cite{curcio2023}, possibly due to the fact that these bands are only visible at specific $k_z$ values (Supplementary section A). Additionally, they cannot simply be the product of current-induced spectral broadening, as minimal broadening with respect to the equilibrium S phase is observed in the S* phase (where the applied voltage is even greater than in the L* phase as seen in Fig.~C.1)


Next, we compare our transport-ARPES results to DMFT calculations (Fig.~\ref{fig3}f - j). As the band disperses from $X$ to $M$ in the insulating S phase (Fig.~\ref{fig3}g), it flattens out before reaching the Fermi level, consistent with experimental observations (Fig.~\ref{fig3}b) and other DMFT calculations~\cite{gorelov2010}. In the L* phase (Fig.~\ref{fig3}h), the band is seen to cross the Fermi level before flattening in the unoccupied region of the spectrum around the $M$ point. Although unoccupied states cannot be resolved with standard ARPES, the depletion of spectral weight experimentally observed at $k^*$ is consistent with the scenario sketched by DMFT. Aside from the difference in distribution of spectral weight between the lower and upper Ru bands, the calculated S phase MDC (Fig.~\ref{fig3}f) agrees well with the experiment. While the agreement seems poor for the L* phase, the increase in intensity at $X$ is still present. The cascade of calculated EDCs (Fig.~\ref{fig3}i, j) also shows the modulation of spectral weight across the $MXM$ cut in the S phase and the momentum independent shape of the spectral weight in the L* phase, similar to that observed in the experiment (Fig.~\ref{fig3}d, e).

\begin{figure*}
    \centering
    \includegraphics[width=0.999\textwidth]{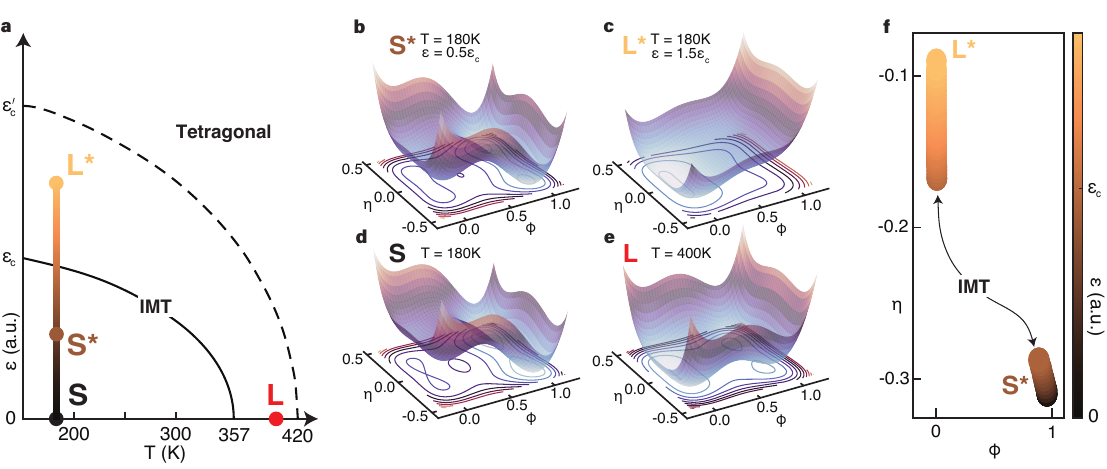}
    \caption{\textbf{Phase diagram and free energies.} (a) Phase diagram as obtained by Landau theory with the first- (solid) and second-order (dashed) phase boundaries; the gradient line corresponds to the values of electric field in (f) and the circles mark the parameters of the free energies plotted in (b - e). (b-e) Free energies for different values of temperature and electric field with local and global minima highlighted by the contour plots. (f) Evolution of the global energy minimum in the space of order parameters $\phi$ and $\eta$, as a function of electric field $\varepsilon$; at critical field $\varepsilon_c$, there is a first order phase transition from the insulating S* ($\phi \approx 1$) to the metallic L* phase ($\phi = 0$).}
    \label{fig:free_energy}
\end{figure*}


We then measured the dispersion at the perpendicular Brillouin zone edge, along $YM$. Data displayed in Fig.~\ref{fig3} are measured with the analyser slit along the crystallographic $b$-axis; to observe the $YM$ cut, we rotate the crystal by 90$^\circ$ to align the analyser slit along the $a$-axis. This allows the photoemission matrix element to access different orbital symmetries, while maintaining $\sigma$-polarized light, resulting in the Fermi surface displayed in the top half of Fig.~\ref{fig4}g (the bottom half of the figure is taken from Fig.~\ref{fig2}a to present the complete polarization-dependent L* Fermi surface for comparison). Surprisingly, no significant dispersion is found along the $YM$ cut, as shown in Fig.~\ref{fig4}b, with spectral weight being most concentrated at the $Y$ and $M$ points. Comparing Figs.~\ref{fig3}b with \ref{fig4}b, we see the S phase is quite anisotropic between the $XM$ and $YM$ directions. However, the $XM$ and $YM$ dispersions computed by DMFT are nearly indistinguishable despite including the orthorhombic asymmetry (Supplementary section D), suggesting that underlying electronic correlation effects already exist in the zero-current low-temperature Mott phase, that have yet to be accounted for in the calculations.

Intriguingly, no visible modification of the Ru dispersion occurs along $YM$ across the IMT (Fig.~\ref{fig4}b, c), aside from a redistribution of spectral weight from the higher to lower energy Ru-band. This is highlighted by the consistent shape of the MDCs (Fig.~\ref{fig4}a) and EDCs (Fig.~\ref{fig4}d, e), in striking contrast to the changes observed along the $XM$ direction (Fig.~\ref{fig3}a - e). This contrast in the electronic structure evolution across the IMT along the two high-symmetry directions can be attributed to a smaller corresponding change in lattice constant along the $a$-axis, $\Delta a_{\mathrm{(S - L^*)}}$ = 0.04~\AA, as compared to the $b$-axis, $\Delta b_{\mathrm{(S - L^*)}}$ = 0.2~\AA ~\citep{bertinshaw2019}. In fact, the measured $XM$ dispersion in the L* phase (Fig.~\ref{fig3}c) more closely resembles the $YM$ dispersion (Fig.~\ref{fig4}c); in other words, the L* phase is more isotropic than the S phase. Therefore, our results suggest that the decreased orthorhombicity strongly influences the band structure evolution, in agreement with previous current-induced \citep{bertinshaw2019} and strain \citep{ricco2018} works on the IMT, but with further electronic correlations at play, which we discuss below.

Let us first comment on the role of Joule heating in this current-induced IMT, which is heavily discussed in the literature \cite{okazaki2013, fuersich2019, mattoni2020, voltage_induced}. To gain further insight, particularly from an electronic structure perspective, we compare the transport-ARPES data in the current-induced metallic L* phase for both analyser slit orientations (Fig.~\ref{fig4}g) with zero-current ARPES data on the high-temperature metallic L phase (Fig.~\ref{fig4}h). All experimental conditions (i.e. light polarization, photon energy, etc.) remain the same between the two measurements, aside from the temperature ($T$ = 180~K and 384~K, respectively). While the Fermi surface characteristics of the L* phase exhibit a clear dependence on experimental geometry, the L phase Fermi surface shows the same spectral features (within the limits of thermal broadening), when measured along both in-plane axes (Fig.~F.1). This indicates a much smaller electronic anisotropy at high temperatures, in accordance with the further reduced orthorhombicity of the L phase compared with the L* phase~\cite{friedt2001}. The differences in the Fermi surfaces are further highlighted by the MDCs taken at the edge of the BZ (Fig.~\ref{fig4}f), where significant intensity is found at the $M$ points for the L* phases (yellow), while intensity peaks in the center of the high-temperature L phase curve and progressively drops off towards the edges of the Brillouin zone (red). These differences confirm that the current-induced L* phase is indeed electronically distinct from the L phase, rather than a consequence of Joule heating.

To gain deeper insight into the nature of the field- versus temperature-induced IMT and the interplay between structural and electronic degrees of freedom, in the following we present a free energy analysis. The IMT at finite temperature and zero electric field is well-understood and described within Landau theory with local lattice distortions as order parameters\cite{caseofcaruo}. A characterization of the free energy in terms of the (correlated) electronic degrees of freedom is equally valid, and can be directly related to the lattice distortions via e.g. DMFT calculations\cite{Georgescu2022}. The electronic order parameter $\phi$ is defined by the disproportion of orbital population of $d_{xy}$ and $d_{xz/yz}$ orbitals\cite{verma2023picosecond}, $\phi = n_{xy} - (n_{xz} + n_{yz})/2$, where $n$ is the expectation value of the respective number operator. For $\phi = 0$ all orbitals are equally populated and the system is metallic, whereas $\phi = 1$ describes the orbitally dependent Mott insulating state with completely filled $d_{xy}$ and half-filled $d_{xz/yz}$ orbitals. Even though applying a voltage and increasing the temperature both result in a first order IMT, our experiment suggests that the L and L* phase are electronically distinct, which raises the question of the underlying mechanism leading to said behaviour. Experimentally, temperature and electric field are alike in that both redistribute the electronic occupation, but different in that the electric field introduces a directionality. We note that since the field drives the system out of equilibrium, the free energy is strictly speaking only well-defined in its absence. We can describe its manifestations effectively, and account for the above mentioned directionality as well as the observed L* orthorhombicity, by introducing a second order parameter $\eta$ that we relate to the imbalance in electron population between the $d_{xz}$ and $d_{yz}$ orbitals, $\eta = \alpha\ (n_{yz} - n_{xz})$, and impose it to be zero in the tetragonal phase, with $\alpha$ as some prefactor.

While determining the exact relation between $\eta$ and the orbital populations would require non-equilibrium DMFT or a similarly elaborate numerical investigation, we are able to capture the field-induced IMT qualitatively using just Landau theory. We expand the free energy $F(\phi, \eta)$ up to fourth power, and couple the two order parameters to the lowest order that is allowed by symmetry (see Supplementary section E for more details). We use the parameters from Ref.\cite{Georgescu2022}, and choose the undetermined coefficients such that: (i) the new phase boundary describes a second order transition to the undistorted tetragonal phase at $420$ K \footnote{Note that while a pure tetragonal phase is found above $420$ K, the crystal is already fairly isotropic in the measured L-phase temperature of $380$ K, hence the lack of dependence on orientation seen in the high temperature Fermi surfaces shown in (Fig.~F.1) \cite{caseofcaruo}.}; and (ii) the expected metallic behaviour of entropy and current density across a first order phase transition is reproduced. Fig.~\ref{fig:free_energy}b - e show $F(\phi, \eta)$ for the different phases with its local and global minima. Tracking the the evolution of the extrema over a range of temperatures and electric field strengths, we produce the phase diagram as shown in Fig.~\ref{fig:free_energy}a. Thermodynamically, the L and L* phase, as well as the S and S* phase, are the same; however, varying values of $\eta$ reflect the different orbital populations over the range of electric fields and temperatures, within that phase region.

\section*{Discussion}\label{sec1}

We have unveiled the current-induced electronic band structure of CRO using the powerful method of transport-ARPES (ARPES conducted in a four-point transport configuration). By establishing core levels as a reference point for energy shifts due to stray electric fields, we have demonstrated the reduction of the gap through the IMT. We have also observed a change in the dispersion of the Ru $t_{2g}$ band, occurring only along the $XM$ direction, i.e. parallel to the $b$-axis and along the direction of current flow. We point out that the current-induced nanostripe pattern only manifests when the current is applied parallel to the $b$-axis \cite{zhang2019}, emphasizing the strong relationship between the DC current and lattice. However, a comparison to state-of-the-art calculations points to a need for future work to further improve the theoretical descriptions of the electronic interaction phenomenology in equilibrium CRO.

With Landau theory calculations we show the lattice must work in concert with an orbitally-driven electronic response to drive the current-induced IMT. In addition, we have proven the current-induced metallic L* phase to be electronically distinct from the temperature-driven metallic L phase, confirming that the former is indeed a unique state. Our results will motivate further theoretical investigations of the current-induced IMT in CRO; this will likely require the explicit inclusion of a non-equilibrium component representing the current.

Beyond the specific case of CRO, this study will spark the investigation of current-induced phase transitions with transport-ARPES in many other strongly correlated materials.

\section*{Methods}\label{sec11}

High-quality single crystals of Ca$_2$RuO$_4$ were grown using the optical floating zone technique as described in Ref.~\cite{nakatsuji2001}. Samples were characterized with x-ray Laue diffraction equipped with a tungsten source and a MWL120 Real-Time Back-Reflection Laue camera, as well as with magnetometry measured in a Quantum design MPMS SQUID to confirm $T_{\mathrm{N}}$ = 110~K and no impurities from additional phases were present. The crystals are shaped with a diamond wire saw to an average dimension of approximately 2 x 1 x 0.25~mm$^{3}$. Sculpting the crystals into a simple rectangular shape is essential to provide the current with a more homogeneous and efficient path. Characterizing ARPES measurements in the zero-current S phase were conducted at the Quantum Materials Spectroscopy Centre beamline at the Canadian Light Source using a horizontal electron analyser slit geometry. The crystals were cleaved \textit{in-situ} at 6 $\times$ 10$^{-11}$~mbar using a ceramic top post. Spectra were measured at 74~eV with $\sigma$- and $\pi$-polarized light at $T = 180$~K to avoid charging.

Transport-ARPES experiments were carried out at the MAESTRO beamline (7.0.2) at the Advanced Light Source at Lawrence Berkeley National Lab using a horizontal electron analyser slit geometry. The crystals were cleaved \textit{in-situ} at 1.7 $\times$ 10$^{-11}$~mbar using a ceramic top post. Spectra were measured at 74~eV with $\sigma$- and $\pi$-polarized light. Current was applied parallel to the $b$-axis in a four-probe configuration using a Keithley 2400 Source Measure Unit. Before the start of measurements, the ground was confirmed to be the same as that of the electron analyser. The samples were kept at $T$ = 180~K under constant cooling to prevent Joule heating. High temperature ARPES experiments were also carried out at MAESTRO at $T$ = 380~K. The manipulator was heated from 10~K, and following heating the manipulator position was allowed to settle for 4~h. Samples were cleaved \textit{in-situ} at 2 $\times$ 10$^{-10}$~mbar using a ceramic top post.

To mount and prepare the samples for transport-ARPES measurements at MAESTRO, conductive EPO-TEK H20E epoxy was used to glue five 75~$\mu$m gold wires to the crystal and attach the wires to the gold plate contacts of the chip carrier. Insulating and thermally efficient Loctite Stycast 2850FT with catalyst 24LV is then used to glue the crystal, wire-side down, onto a sapphire substrate. Both epoxies are thermally efficient and were cured just below 80~$^\circ$C for over 4 hours. The substrate prevents the crystal from making contact with the gold base of the chip carrier. It is also important to glue the contacted face of the crystal to the substrate as the wires must be secure during the crystal cleave. The chip carrier can then be placed on the custom-made transport pucks of the MAESTRO beamline, shown in Fig. 1a, which feature eight primary contacts made of gold-coated beryllium copper and secured with titanium screws and molybdenum nuts. Phosphor bronze clips are used to hold the chip carrier in place. The puck itself is made of oxygen-free copper and has a protruding piece at the center, hugged by the ceramic frame, to ensure strong thermal coupling with the sample. 

Our (single-shot) DFT+DMFT calculations were performed using the crystallographically refined atomic coordinates for the S, S*, L* and L phases, as obtained from neutron diffraction. After convergence of the DFT calculation for each of these structures (using the VASP code \cite{vasp} with a full potential linearized augmented plane-wave method and PBE-GGA exchange-correlation functional), we downfolded the Kohn-Sham wavefunctions to an effective Wannier basis \cite{mostofi2008} covering an energy window of $-2$~eV to $1$~eV around the Fermi level $E_{\mathrm{F}}=0$. The resulting three band model originates from the cubic Ru $t_{2g}$ states and includes: i) non-cubic crystal field distortions, as well as ii) the spin-orbit coupling operator projected to the $t_{2g}$ states. The interaction is approximated by the rotationally invariant Kanamori operator with a Hubbard $U = 1.9$~eV and Hund's coupling $J = 0.4$~eV (following ~\textcite{kim2018}) for all structures. For the solution of the DMFT auxiliary impurity models, we employ the continuous-time hybridization-expansion quantum Monte-Carlo solver (CT-QMC) of the TRIQS library \cite{seth2016, parcollet2015}. After DMFT convergence we performed analytical continuation of the self-energy by maximum-entropy method (MaxEnt) following Kraberger et al. \cite{kraberger2017}. The resulting real-frequency self-energies were then used to produce the orbitally- and momentum-resolved single-particle spectral functions which are shown in this work. Comprehensive technical details and plots (e.g. of the analytically continued self-energies) can be found in the supplemental material of Ref.~\onlinecite{bertinshaw2019}.

\section*{Acknowledgments}
We acknowledge useful discussions with Andrew J. Millis, Jan Bruin, Stephanie Gilbert-Corder, Christopher Guti\'{e}rrez, Sae-Hee Ryu, Henri Menke, Nobumichi Tamura, and Giorgio Levy, as well as MengXing Na and Alex Anees for figure design. This research was undertaken thanks in part to funding from the Max Planck-UBC-UTokyo Centre for Quantum Materials and the Canada First Research Excellence Fund, Quantum Materials and Future Technologies Program. This project is also funded by the Killam, Alfred P. Sloan, and Natural Sciences and Engineering Research Council of Canada’s (NSERC’s) Steacie Memorial Fellowships (A.D.); the Alexander von Humboldt Foundation (A.D.); the Canada Research Chairs Program (A.D.); NSERC, Canada Foundation for Innovation (CFI); the Department of National Defence (DND); the British Columbia Knowledge Development Fund (BCKDF); the Mitacs Accelerate Program; the QuantEmX Program of the Institute for Complex Adaptive Matter (ICAM); the Moore EPiQS Program (A.D.); and the CIFAR Quantum Materials Program (A.D.). Use of Advanced Light Source (MAESTRO beamline) at Lawrence Berkeley National Lab is supported by the U.S. Department of Energy, Office of Science User Facility under Contract No. DE-AC02-05CH11231. Use of the Canadian Light Source (Quantum Materials Spectroscopy Centre), a national research facility of the University of Saskatchewan, is supported by CFI, the NSERC, the National Research Council, the Canadian Institutes of Health Research, the Government of Saskatchewan, and the University of Saskatchewan. S.S. acknowledges support by the Netherlands Organisation for Scientific Research (NWO 019.223EN.014, Rubicon 2022-3). C.T.S. acknowledges support from the NSERC and by an ALS Doctoral Fellowship in Residence. \\

\section*{Conflict of interest}
The authors declare no competing interests.

\section*{Availability of data and materials}
All relevant data are available from the corresponding author upon reasonable request.

\section*{Author contributions}
C.T.S., B.K., E.R. and A.D. conceived the project and designed the experiments. M.K. grew the crystals and C.T.S and P.P. characterized them. C.T.S., I.M, M.Z, N.-H.J., S.Z., S.S, C.A-Y., V.Z., and B.Z. carried out the ARPES experiments with assistance from S.G., C.J., A.B. and E.R.. C.T.S. performed the data analysis and I.M., M.Z., S.Z, N.-H.J., I.S.E, C.J., A.B., E.R., and A.D. provided input and assistance. M.S. and P.H. conducted and analysed the DFT and DMFT calculations. N.H. and M.F. carried out the free energy analysis of the current-induced phase. C.T.S wrote the manuscript with significant input from N.H., I.M., and M.Z. and input from all authors. B.K and A.D. were responsible for the overall direction, planning, and management of the project.

\bibliography{sn-article}

\end{document}